\documentclass[journal,twoside]{IEEEtran}
\usepackage{cite}
\usepackage{amsmath,amssymb,amsfonts}
\usepackage{algorithmic}
\usepackage{graphicx}
\usepackage{textcomp}
\usepackage{upgreek}
\usepackage{cases}
\usepackage{layouts}
\def\BibTeX{{\rm B\kern-.05em{\sc i\kern-.025em b}\kern-.08em
    T\kern-.1667em\lower.7ex\hbox{E}\kern-.125emX}}
\begin{document}
\title{Dual-energy X-ray dark-field material decomposition}
\author{Thorsten Sellerer*, Korbinian Mechlem*, Ruizhi Tang, Kirsten Taphorn,  Franz Pfeiffer, Julia Herzen
\thanks{We  acknowledge  financial  support  through  the  DFG  (Gottfried  Wilhelm  Leibniz  program),  the  European  Research Council (AdG 695045) and the DFG (Research Training Group GRK 2274).}
\thanks{K. Taphorn and  F. Pfeiffer are with the Department of Diagnostic and Interventional Radiology, Klinikum rechts der Isar, Technical University of Munich.}
\thanks{K. Mechlem, T. Sellerer, R. Tang,  K. Taphorn, J. Herzen  and F. Pfeiffer are with the Chair of Bio Physics, Department of Physics and Munich School of BioEngineering, Technical University of Munich, 85748 Garching, Germany.}
\thanks{F. Pfeiffer is with the Institute for Advanced Study, Technical University of Munich, 85748 Garching, Germany.}
\thanks{\textbf{* T. Sellerer and K. Mechlem are co-first authors (email: thorsten.sellerer@tum.de; korbinian.mechlem@ph.tum.de).}}}

\maketitle
\begin{abstract}
Dual-energy imaging is a clinically well-established technique that offers several advantages over conventional X-ray imaging. By performing measurements with two distinct X-ray spectra, differences in energy-dependent attenuation are exploited to obtain material-specific information. This information is used in various  imaging applications to improve clinical diagnosis. In recent years, grating-based X-ray dark-field imaging has received increasing attention in the imaging community. The X-ray dark-field signal originates from ultra small-angle scattering within an object and thus provides information about the microstructure far below the spatial resolution of the imaging system. This property has led to a number of promising future imaging applications that are currently being investigated. 
However, different microstructures can hardly be distinguished with current X-ray dark-field imaging techniques, since the detected dark-field signal only represents the total amount of ultra small angle scattering.
To overcome these limitations, we present a novel concept called dual-energy X-ray dark-field material decomposition, which transfers the basic material decomposition approach from attenuation-based dual-energy imaging to the dark-field imaging modality. We develop a physical model and algorithms for dual-energy dark-field material decomposition and evaluate the proposed concept in experimental measurements. Our results suggest that by sampling the energy-dependent dark-field signal with two different X-ray spectra, a decomposition into two different microstructured materials is possible. Similar to dual-energy imaging, the additional microstructure-specific information could be useful for clinical diagnosis.
\end{abstract}


\section{Introduction}
\label{sec:introduction}

\IEEEPARstart{T}{he} 
concept of dual-energy imaging, which was firstly introduced in 1976 \cite{Alvarez1976},  represents a great milestone in diagnostic X-ray imaging. Due to its capability to extract material-specific information non-invasively from an object, dual-energy imaging initiated the development of several clinical applications. By performing measurements with two distinct X-ray spectra, differences in energy-dependent attenuation are exploited to distinguish between different materials. In the last years, dual-energy computed tomography (DECT) has become a powerful and well-established diagnostic tool in the daily clinical workflow. DECT has turned out to be particularly valuable for abdominal imaging, where it has led to significant improvement in diagnostic imaging. The availability of additional information in the form of  virtual monochromatic images and quantitative material-specific contrast agent density maps can provide an improved detectability of oncological \cite{Agrawal2014, Marin2014} and vascular \cite{Pinho2013, Flors2013} pathologies. 
Apart from DECT, there are several dual-energy applications relying on projection data only, such as dual-energy X-ray absorptiometry \cite{Laskey1996}, dual-energy subtraction radiography \cite{Vock2009} and contrast-enhanced digital mammography \cite{Badr2014}. 

Grating-based differential phase-contrast (DPC) \cite{Pfeiffer2006} and dark-field \cite{Pfeiffer2008} imaging are two emerging imaging modalities that use entirely different contrast generation mechanisms compared to attenuation-based techniques. In phase-contrast imaging, the signal originates from the phase-shift an X-ray wave-front undergoes when penetrating an object. As this phase-shift is not directly accessible, a three grating interferometer is utilized to measure the refraction angle that is associated with the induced phase-shift. Although DPC imaging can achieve a highly improved soft-tissue contrast compared to attenuation-based imaging \cite{Herzen2009, Zambelli2010, Donath2010, Chen2011, Engel2011}, the benefits strongly depends on the imaging parameters \cite{Raupach2011, Raupach2012, Mechlem2020_2}. 
Besides the differential phase-contrast signal, the aforementioned interferometer can also be used to extract the X-ray dark-field signal \cite{Pfeiffer2008}, which is related to ultra-small angle scattering within the object and provides information about the microstructure far below the resolution of the imaging system.
Following evidence of improved diagnosis and staging of emphysema in mice \cite{Schleede2012}, this novel imaging modality has attracted increasing attention in the  imaging community. Many pre-clinical studies have focused on lung imaging since the microstructured air-tissue interfaces generates a strong dark-field signal. Dark-field radiography has the potential to improve the detection of emphysema \cite{hellbach2015vivo}, pulmonary fibrosis \cite{yaroshenko2015improved} and chronic obstructive pulmonary disease \cite{Willer2018}.  Other potential clinical applications of X-ray dark-field imaging include foreign body detection \cite{braig2018simultaneous}, mammography \cite{grandl2015improved,scherer2016improved} and  contrast agent imaging \cite{velroyen2013microbubbles}.\\
Significant progress has recently been made in transferring dark-field radiography from small animal research to clinically relevant imaging parameters, particularly with regard to sample size, field of view and X-ray energy.
The technique has been successfully demonstrated in in-vivo pig \cite{Gromann2017} and ex-vivo human cadaver studies \cite{Willer2018}.\\
In order to overcome the aforementioned limitations of conventional grating-based X-ray dark-field imaging, we introduce a novel concept called dual-energy X-ray dark-field material decomposition. Contrary to attenuation-based imaging, the dark-field signal is not only influenced by the chemical composition of an object (via the electron density), but also strongly depends on the properties of the microstructure \cite{Prade2015, Yashiro2010}.
Similarly to the attenuation of X-rays, the dark-field signal has a distinct energy dependency which has been shown to vary for materials with a different microstructure \cite{Taphorn2020}.
In this work we aim at establishing a link between the structural properties of an object and the dependency of the corresponding dark-field signal on the photon energy. Based on these considerations, we transfer the basis material decomposition approach from attenuation-based dual-energy imaging to the dark-field imaging modality. Finally, we validate the feasibility of the proposed method by conducting several proof-of-concept experiments.

\section{Methods}
\subsection{Spectral phase-contrast and dark-field model}
\label{Spectral Phase-contrast and Darkfield Model}

\begin{figure*}[t!]
    \centering
    \includegraphics[width=\textwidth]{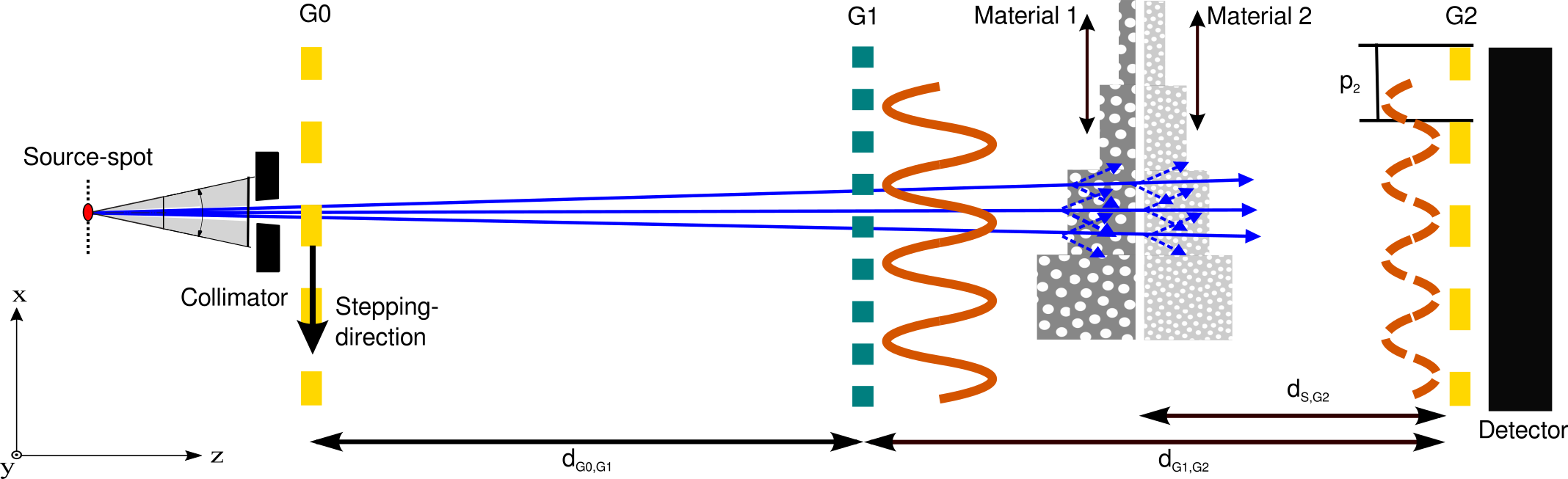}
    \caption{Experimental set-up. The experimental measurements were conducted on a lab-bench set-up with a microfocus tube, a symmetric Talbot-Lau interferometer and a flat-panel detector. For the calibration procedure, different thicknesses of the basis materials were mounted onto movable linear-stages.}
    \label{setup}
\end{figure*}

In conventional differential phase-contrast and dark-field imaging, a so-called stepping curve is acquired using a three grating interferometer (cf. fig.~\ref{setup}). Thereby, several measurements with slightly transversally shifted positions of one of the gratings are acquired. The expected intensity for each stepping position is commonly approximated by \cite{Teuffenbach2017}:

\begin{equation}
    \label{conv_dpc}
    \hat{y}_i^r = Se^{-\mu d^\mu_i} \left(1+ Ve^{-\epsilon d^\epsilon_i}\cos{(\phi^{r}_i + \Delta\phi_i}) \right),
\end{equation}
where $\hat{y}_i^r$ is the expected intensity for pixel $i$ and stepping position $r$. The quantity $d^\mu_i$ describes the thickness of an attenuating material in the beam path and $\mu$ its corresponding linear attenuation coefficient. The visibility parameter $V$ represents the ratio of the amplitude and the mean value of the sinusoidal stepping curve. In general, the visibility can be defined by the maximum and minimum intensity values of an intensity modulation:
\begin{equation}
V = \frac{ \hat{y}_{\mathrm{max}} - \hat{y}_{\mathrm{min}}} {\hat{y}_{\mathrm{max}} + \hat{y}_{\mathrm{min}}}.
\end{equation}
Ultra-small angle X-ray scattering on a microstructured object reduces the visibility of the stepping curve and thus generates a dark-field signal. Similarly to the Lambert-Beer law for the attenuation channel, the visibility reduction compared to the reference visibility $V$ (without the object in the beam path) is described by an exponential term $\left(e^{-\epsilon d^\epsilon_i}\right)$. 
 The thickness of the dark-field scatterer is given by $d^\epsilon_i$ and $\epsilon$ is the material's linear diffusion coefficient, which can be defined analogously to the linear attenuation coefficient $\mu$ \cite{Bech2010}. The displacement of the stepping curve due to the refraction caused by the sample is labeled as $\Delta\phi$. The reference intensity of the incident beam is given by $S$ and $\phi^r$ describes the reference phase of the stepping curve for stepping position $r$. The standard stepping curve model of eq. \ref{conv_dpc} neglects beam hardening effects by implicitly assuming that the polychromatic X-ray spectrum can be described by an effective energy $E_{\rm{eff}}$:
\begin{equation}
S = \int_0^{E_{\mathrm{V}}} S(E), \ \mu = \mu(E_{\rm{eff}}),  \ \epsilon = \epsilon(E_{\rm{eff}}),
\end{equation}
where $S(E)$ is the effective X-ray spectrum seen by the detector and $E_{\rm{V}}$ denotes the maximum photon energy given by the acceleration voltage of the X-ray tube.
In the context of grating-based spectral differential phase-contrast X-ray imaging, we have recently extended the standard stepping curve model to include polychromatic effects \cite{Mechlem2020_1}:
\begin{equation}
    \label{sdpc}
    \begin{split}
    \hat{y}_i^{rs} = &\int^{E_{\mathrm{V}}}_0   S^s(E) e^{-\mu_1(E) d^{\mu_1}_i - \mu_2(E)d^{\mu_2}_i} \\\
    & \left[1+ V(E)e^{-\epsilon(E) d^\epsilon_i}\cos{(\phi^{r}_i + \Delta\phi_i(E))} \right].
    \end{split},
\end{equation}
Compared with eq. \ref{conv_dpc}, an additional index $s$ is introduced to indicate the different X-ray spectra used for spectral imaging.
The quantity $\hat{y}_i^{rs}$ thus represents the expected intensity for the $s$-th effective X-ray spectrum (including source and detector effects), detector pixel $i$ and stepping position $r$.
Excluding the presence of K-edge discontinuities within the relevant energy range, the two material approximation \cite{Alvarez1976} is used to model the attenuation of the incident spectra, where $d^{\mu_1}_i, d^{\mu_2}_i $ are the line integrals of the attenuating basis materials in pixel $i$ and $\mu_1(E), \mu_2(E)$ describe the energy dependencies of the corresponding linear attenuation coefficients. The visibility of the stepping curve becomes an energy-dependent quantity labeled as the visibility spectrum $V(E)$. The visibility reduction  attributed to the dark-field signal is modeled by the thickness of the scattering material $d^\epsilon_i$ and the energy-dependency of the scattering processes $\epsilon(E)$.

A general mathematical description of the dark-field signal in grating-based neutron and X-ray imaging allows to relate the visibility reduction caused by small-angle scattering to the projected autocorrelation function $G(x)$ \cite{Strobl2015}:
\begin{equation}
    \label{DF-signal}
    V_{\mathrm{sca}}(\xi) / {V(\xi)} = \exp{ \left(- d^\epsilon \sigma (E) (1-G(\xi) \right)},
\end {equation}
where $V_{\mathrm{sca}}$ and $V$ describe the visibility with and without scattering structures along the beam path, respectively. The penetrated object thickness is labelled by $d^\epsilon$ and the total scattering cross-section $\sigma(E)$ denotes the scattering probability and $\xi$ is the correlation length of the grating inteferometer. It describes the length at which correlations within the sample are measured \cite{Prade2015} and can be directly related to the the geometrical parameters of the grating interferometer. Assuming that the sample is located between the $G1$ and $G2$ grating (compare figure \ref{setup}), the correlation length is given by:
\begin{equation}
    \label{corr-length}
    \xi = \frac{hc}{E} \frac{d_{S,G2}}{p_2},
\end{equation}
where $h$ and $c$ are Planck's constant and the speed of light, respectively. The distance between the sample and the $G2$ grating is denoted by $d_{S,G2}$ and $p_2$ is the period of the $G2$ grating. 
In the following, we assume that the thickness of the sample is small compared with the inter-grating distances. In this case, $d_{S,G2}$ can be regarded as a constant, which means that $\xi$ and thus also the dark-field signal $V_{\mathrm{sca}}(\xi)/V(\xi)$ in eq. \ref{DF-signal} only depend on the photon energy $E$. \\
We define the cartesian coordinate system $(x,y,z)$ such that the $z$-coordinate coincides with the beam direction. The grating bars are aligned along the $y$ direction, which means that the grating interferometer is sensitive to phase shifts and small angle scattering along the $x$-direction (i.e. perpendicular to the grating bars). With this definition, the projected autocorrelation function along the beam direction is expressed as \cite{Krouglov2003,Andersson2008}:
\begin{equation}
    \label{corr-func}
    G(x) = \frac{1}{\gamma_0} \int \gamma (x,0,z) dz,
\end{equation}
where $\gamma_0$ is a normalization factor such that $G(0) = 1$ and $\gamma (x,y,z) = \gamma(\vec{r})$ is the normalized autocorrelation function of the electron density fluctuations of the sample:
\begin{equation}
\begin{split}
&\gamma(\vec{r}) = \frac{\int \Delta \rho_{\mathrm{el}}(\vec{r\,}') \Delta \rho_{\mathrm{el}}(\vec{r\,}'+ \vec{r}) d\vec{r\,}'}{\int \Delta \rho_{\mathrm{el}}(\vec{r\,}') \Delta \rho_{\mathrm{el}}(\vec{r\,}') d\vec{r\,}'}, \\
&\Delta \rho_{\mathrm{el}}(\vec{r}) = \rho_{\mathrm{el}}(\vec{r}) - \left< \rho_{\mathrm{el}}(\vec{r}) \right>.
\end{split}
\end{equation}
Note that in case of using a standard grating interferometer the real-space auto-correlation function can only be sampled in the $x$-direction, i.e perpendicular to the grating bars. 

\begin{table*}[t]
\caption{Grating specifications and acquisition parameters for the experimental measurements.}
\label{specs}
\fontsize{10}{12}\selectfont
\centering
\begin{tabular}{c c c c c c}
  \hline
  \hline
  Grating & Effect & Material & Height [$\upmu\text{m}$] & Period [$\upmu\text{m}$] & Duty cycle\\ 
  \hline
  $G0$ & absorption & Gold & $160-170$ & $10$ &  $0.5$ \\
  $G1$ & $\pi/2$ phase-shift & Nickel & $8$ & $5$ &  $0.5$ \\
  $G2$ & absorption & Gold & $160-170$ & $10$ & $0.5$ \\
  &&&&& \\
  \hline
  \hline
  Design energy $E_D$ [keV] & $d_{G0,G1}\,\text{[cm]}$ & $d_{G1,G2}\,\text{[cm]}$ & $d_{S,G2}\,\text{[cm]}$ & $\xi(E_D) \ [\upmu\text{m}$] & Eff. pixel size [$\upmu\text{m}$] \\ 
  \hline
  $45$ & $92.5$ & $92.5$ & $46.25$ & $1.27$ & $150$\\
  &&&&& \\
  \hline
  \hline
  Source spectrum & Tube voltage [kV] & Tube power [W] & Filter & Phase steps & Time per step [s] \\ 
  \hline
  Low & $50$ & $100$ & - &  7 & 5  \\
  High & $80$ & $180$ & $0.25\,\text{mm}$ Mo &  7 & 5  \\
\end{tabular}
\end{table*}

According to eq.~\eqref{DF-signal}, the energy-dependency of the dark-field signal can be divided into two constituents: The energy-dependency of the scattering cross-section $\sigma(E) \sim 1/E^2$ and the energy-dependency introduced by probing the samples (projected) autocorrelation with the correlation-length $\xi(E)$, which in turn is a function of the photon energy. The scattering cross-section $\sigma(E)$ only depends on the average electron density within an image voxel (or pixel) and is thus independent of the microstructure. However, the energy-dependency introduced by the correlation function $G(\xi)$ is specific to  the micro-structural properties of the object. As an example, we consider a diluted suspension of monodisperse microspheres. In this case, the correlation function $G(\xi)$ can be well approximated by \cite{Andersson2008}:
\begin{equation}
    \label{corr-function-sphere}
    G_{sphere}(\xi) \approx \exp \left(-\frac{9}{8} \left(\frac{\xi}{R}\right)^2\right) = \exp \left(-\frac{9}{8} q^2\right),
\end{equation}
where the dimensionless parameter $q=\xi /R$ gives the ratio of the correlation length to the radius $R$ of the spheres.

\begin{figure}
    \centering
    \includegraphics{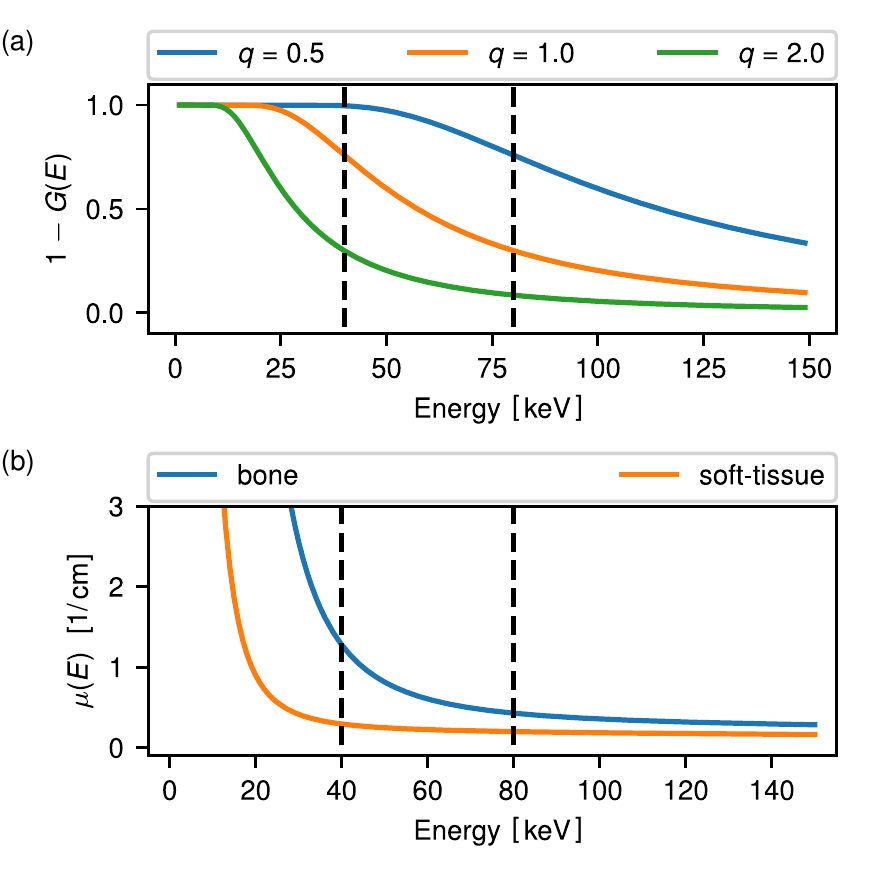}
    \caption{Analogy between the dark-field and the attenuation signal. The correlation-function plotted against the photon energy \textbf{(a)} varies depending on the ratio $q$ between the size of a sample's microscopic structure $R$ and the correlation-length $\xi(E_D)$ at the design energy of the grating interferometer. This behavior is analogous to the variation of the energy-dependency of the attenuation signal caused by materials with different chemical compositions \textbf{(b)}.}
    \label{fig:att_vs_df}
\end{figure}

For a better visualization, fig.~\ref{fig:att_vs_df}(a) shows $1-G_{sphere}(\xi)$ as a function of the photon energy for three different sphere sizes. As illustrated in the figure, the energy-dependency of the correlation-function varies for different ratios of the correlation length and the sphere size. This behavior is analogous to the varying energy-dependency of the attenuation caused by materials with different atomic numbers (cf. fig.~\ref{fig:att_vs_df}(b)).

By performing a limit value consideration for sphere sizes far larger and smaller than the correlation length of the grating interferometer, we arrive at:
\begin{numcases}{G(\xi) \approx}
\label{limit1}
0,  & for \  $q \gg 1$\\
\label{limit2}
1- \frac{9}{8} \left(\frac{\xi}{R}\right)^2, & for \ $q < 1$.
\end{numcases}

Finally inserting eq.~\eqref{limit1} and eq.~\eqref{limit2} into eq.~\eqref{DF-signal} and assuming approximately equally sized spheres, we obtain:
\begin{numcases}{\epsilon(E)  \sim}
\label{limit21}
E^{-2},  & for  \ $q \gg 1$ \ \\
\label{limit22}
E^{-4}, & for \ $q < 1$, \
\end{numcases}
revealing a distinct variation in the energy-dependency of the dark-field signal for differently sized microspheres (compared with the correlation length of the interferometer). Although the considerations above only apply to samples consisting of a homogeneous distribution of equally sized microspheres, a generalization of the concept to arbitrary microstructures is possible. This is due to the general properties of the correlation function:

\begin{numcases}{G(x)=}
\label{limit31}
0, & for \ $q \gg 1$\\
\label{limit32}
]-1,1[ & for \ $q \sim 1 $ \ \\
\label{limit33}
1,  & for \ $q \ll 1$,
\end{numcases}
where the parameter $q$ is generalized to the ratio of the correlation length and the typical structural size of the relevant microstructure. Equations \eqref{limit31}-\eqref{limit33} apply for microstructures that have no long-range order. 
Consequently, the correlation-length of a Talbot-Lau interferometer can normally be chosen in such a way that two  differently sized microstructures within an object show a difference in the energy-dependency of the induced dark-field signal and therefore allow for a differentiation based on energy-resolved measurements. Grounded on these considerations we are able to set up the following system of equations:
\begin{equation}
    \label{DF-decomp}
    \begin{split}
    -\ln{(V_{\mathrm{sca}}(E_1)/V(E_1))} &= \epsilon_1(E_1) d^{\epsilon_1} +  \epsilon_2(E_1) d^{\epsilon_2} \\
    -\ln({V_{\mathrm{sca}}(E_2)/V(E_2))} &= \epsilon_1(E_2) d^{\epsilon_1} +  \epsilon_2(E_2) d^{\epsilon_2},
    \end{split}
\end{equation}
where $\epsilon_{1}(E)$ and $\epsilon_{2}(E)$  are the energy dependencies of two different dark-field inducing microstructures and the logarithmic visibility fractions respresent measurements with two different photon energies $E_1,E_2$. In analogy to attenuation-based dual-energy material decomposition \cite{Alvarez1976}, eq.~\eqref{DF-decomp} can be solved for the dark-field basis material thicknesses $d^{\epsilon_1}$ and $d^{\epsilon_2}$. Finally, we extend the concept for the use with polychromatic photon spectra by inserting eq.~\eqref{DF-decomp} into eq.~\eqref{sdpc} yielding a generalized form of the previously introduced spectral phase-contrast and dark-field model:
\begin{equation}
    \label{sdpcdf}
    \begin{split}
    \hat{y}_i^{rs} = &\int^{E_{\mathrm{V}}}_0   S^s(E) e^{-\mu_1(E) d^{\mu_1}_i - \mu_2(E)d^{\mu_2}_i} \\\
    & (1+ V(E)e^{-\epsilon_1(E) d^{\epsilon_1}_i -  \epsilon_2(E) d^{\epsilon_2}_i}\cos{(\phi^{r}_i + \Delta\phi_i(E)})).
    \end{split}
\end{equation}
Acquiring at least three different stepping positions with two different photon spectra allows for a pixel-wise decomposition of the measurements into basis material line integrals by maximum-likelihood estimation. Assuming Poisson statistics for the measured photon counts, this corresponds to minimizing the negative log-likelihood function:
\begin{equation}
    \label{log-like}
     -L(d^{\mu_1}_i, d^{\mu_2}_i,  d^{\epsilon_1}_i, d^{\epsilon_2}_i, \Delta\phi_i) = \sum _{r=1}^R \sum _{s=1}^S \hat {y}_{i}^{rs} - y_{i}^{rs} \text {ln}\left ({\hat {y}_{i}^{rs} }\right)),
\end{equation}
where $R$ and $S$ are the number of phase steps and photon spectra, respectively. The number of photons measured in detector pixel $i$ at stepping position $r$  with photon spectrum $s$ is denoted by ${y}_i^{rs}$. In the following, we focus on two distinct energy spectra (dual-energy imaging, $S=2$), however similar to attenuation-based spectral imaging, a generalization of the model to more than two spectra is straightforward. As we have shown in previous work \cite{Mechlem2020_1}, the phase-shift $\Delta\phi^i$ can be eliminated as additional optimization variable by expressing it with the attenuation basis material thicknesses $d^{\mu_1}_i, d^{\mu_2}_i$ via the projected electron density gradient. However, if one is primarily interested in the dark-field channels, regarding $\Delta\phi^i$ as additional optimization variable makes the log-likelihood function  separable with respect to different detector pixels, which greatly accelerates the numerical optimization. Theoretical noise considerations \cite{Mechlem2020_2} show that the dark-field channels are only weakly coupled with the other optimization variables (for a standard phase stepping scan).  Consequently,  making the model separable only leads to a negligible increase of  the noise level for the dark-field images.

\begin{figure*}[t]
    \centering
    \includegraphics[width=\textwidth]{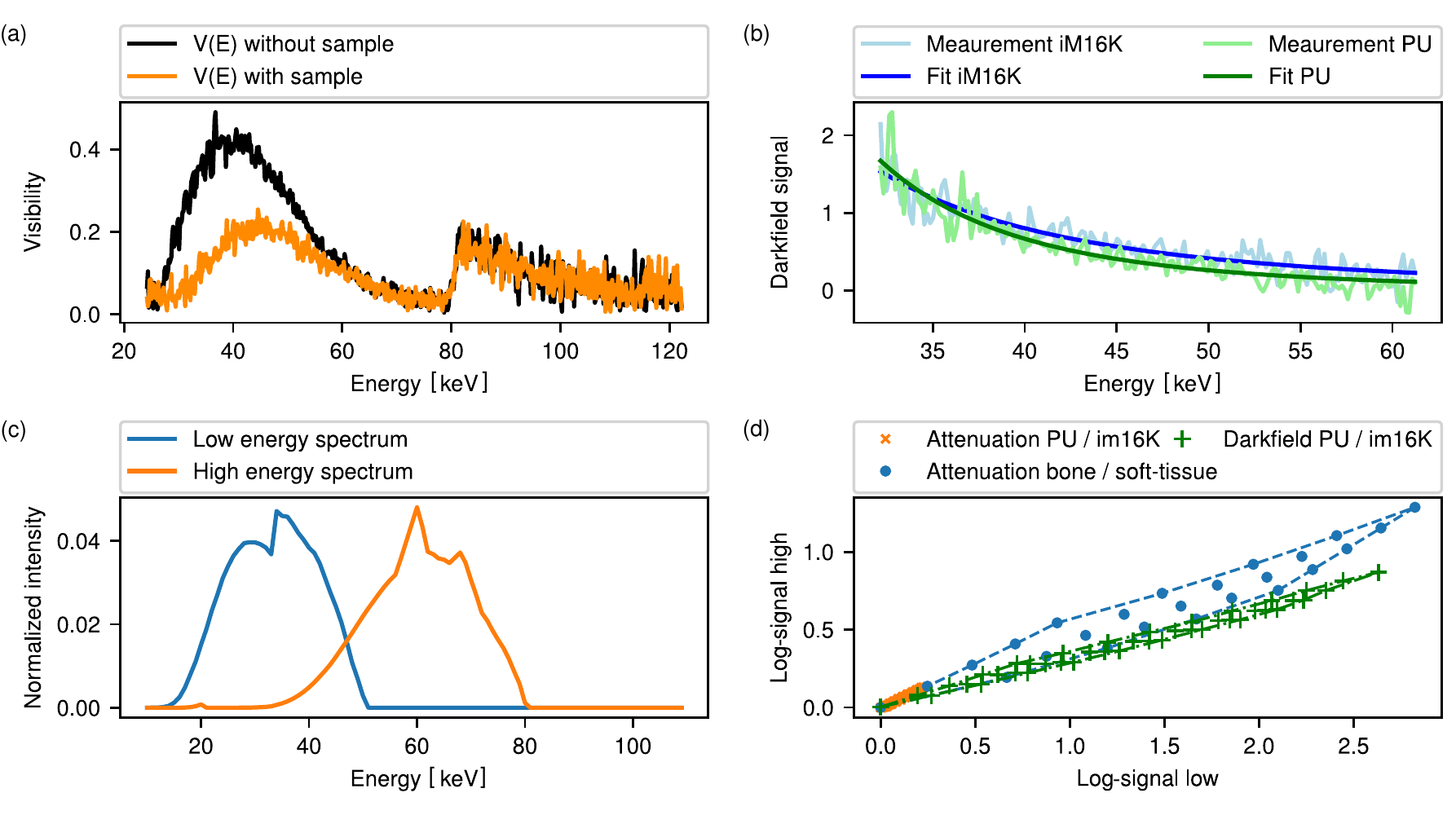}
    \caption{Energy-dependency of the dark-field signal. The energy-dependency of the dark-field signal of different materials was determined by spectroscopic measurements of the visibility spectra \textbf{(a)} with and without the sample in the beam path (exemplarily displayed for the PU sample). The resulting dark-field signals of im16K and PU \textbf{(b)} reveal a different energy-dependency of the two materials. Although the dark-field measurements of the two materials were performed with two well separated photon spectra (the simulated spectra are shown in \textbf{c}), the ratio of the measured log-signals differs only slightly between the two dark-field basis materials (\textbf{d}). For a better comparability with attenuation-based dual-energy imaging, the simulated attenuation log-signals (using bone and soft tissue as basis materials) are also displayed in \textbf{(d)}}.
    \label{fig:e-dep-df}
\end{figure*}
\subsection{Empirical model for dark-field material decomposition}
\label{sec:Empirical Model for Darkfield Material Decomposition}

\begin{table*}
\caption{Investigated materials and corresponding energy-dependency of the dark-field signal. The exponent refers to the mean value of the exponents determined in three independent measurements. The uncertainty of the listed values gives the standard deviation of the obtained exponents.}
\label{mats}
\fontsize{10}{12}\selectfont
\centering
\begin{tabular}{c c c c}
  \hline
  \hline
  Material & Sample thickness [mm] & Average structure size  [$\upmu\text{m}$] & Fitted exponent $x$  \\
  \hline
  PU & $10$  & 200-500 & $-3.89 \pm 0.33$\\
  im16K & $2.5$ & 20 & $-2.70 \pm 0.31$\\
\end{tabular}
\end{table*}

In this work, we focus on a proof-of-concept study for dark-field material decomposition with microstructured samples that show weak attenuation and small phase shifts. In this case, a simplified forward model for the expected visibilities $\hat{V}_{i,\mathrm{sca}}^{s}$ of the stepping curves can be employed:

\begin{equation}
\label{vis_reduction_with_spectrum}
\hat{V}_{i,\mathrm{sca}}^{s}(d^{\epsilon_1}_i,d^{\epsilon_1}_i)  = \int_0^{E_{\mathrm{V}}} S^s_{\mathrm{N}}(E) V(E) e^{-\epsilon_1(E) d^{\epsilon_1}_i -  \epsilon_2(E) d^{\epsilon_2}_i},
\end{equation}

where $S^s_{\mathrm{N}}(E)$ is the normalized effective spectrum, i.e. $\int_0^{E_{\mathrm{V}}} S^s_{\mathrm{N}}(E) =1$.
Knowledge of the photon spectra $S^s(E)$, the visibility spectrum $V(E)$ and the energy-dependencies of the scattering materials $\epsilon_1(E), \epsilon_2(E)$  allows to calculate the dark-field basis material line integrals $d^{\epsilon_1}, d^{\epsilon_2}$ by comparing the forward model with the measured visibilities $V_i^s$.
However, an accurate  experimental determination of these quantities is challenging. In the last years, several empirical forward models that are tuned by calibration measurements have been developed for attenuation-based spectral X-ray imaging. This strategy circumvents the problem of determining the source spectrum and the detector response, which can be particularly challenging for photon-counting detectors. 
Due to the mathematical similarity between the visibility reduction in eq. \ref{vis_reduction_with_spectrum} and the polychromatic Lambert-Beer law, it is possible to adapt an existing empirical polynomial-based forward model \cite{Mechlem2018} (originally developed for attenuation-based spectral imaging) for dark-field material decomposition. Thereby, the visibility reduction measured for photon spectrum $s$ in pixel $i$ is modeled by:
\begin{equation}
\label{fwd-emp}
\hat{V}_{i,\mathrm{sca}}^{s} / {V_{i}^{s}} =  \exp {\left( -P_{i}^{s}( {d}^{\epsilon_1}_{i}, {d}^{\epsilon_2}_{i};\vec {{c}^{s}_{i}}) \right)}, 
\end{equation}
where $P_{i}^{s}({d}^{\epsilon_1}_{i}, {d}^{\epsilon_2}_{i})$ is a second order rational function of the basis material thicknesses with the fit coefficients $\vec{c\,}^{s}_{i} = (c_0^s, ..., c_7^s)^T$:
\begin{equation}
\label{poly}
\begin{split}
&P({d}^{\epsilon_1}_{i}, {d}^{\epsilon_2}_{i})= \\ 
\\
& \frac{c_{0} + c_{1}{d}^{\epsilon_1}_{i} + c_{2}{d}^{\epsilon_2}_{i} + c_{3}({d}^{\epsilon_1}_{i})^2 + c_{4}({d}^{\epsilon_2}_{i})^2 + c_{5}{d}^{\epsilon_1}_{i}{d}^{\epsilon_2}_{i}}{1 + c_{6}{d}^{\epsilon_1}_{i} + c_{7}{d}^{\epsilon_2}_{i}},
\end{split}
\end{equation} 
where the indices $s$ and $i$ of the coefficients $\vec{c\,}^{s}_{i}$ have been omitted for convenience. The fit parameters $\vec{c\,}^{s}_{i}$ are determined by a least-square fit to the calibration measurements:
\begin{equation}
\label{cal-poly}
\vec {c\,}_i^s = \text {arg min} \sum _{k=1}^{K} w_{k} {\left({l_{ik}^{s} - P \left({d}^{\epsilon_1}_{i}, {d}^{\epsilon_2}_{i}; \vec {c\,}_i^s \right) }\right)}^{2},
\end{equation} 
where $l_{ik}^{s} = -\ln \left({V}_{ik,\mathrm{sca}}^{s} / V_i^s \right)$ is the negative logarithm of the visibility reduction measured for calibration point $k$ and the weights $w_k$ consider the statistical uncertainty of the corresponding measurements. After determining the parameters $\vec{c\,}^{s}_{i}$, the calibrated empirical forward model (cf. eq.~\eqref{fwd-emp}) can be used to decompose dual-energy visibility measurements of a sample into the dark-field basis material line integrals ${d}^{\epsilon_1}_{i},{d}^{\epsilon_2}_{i}$. Assuming a Gaussian noise  distribution for the measured visibilities, this corresponds to minimizing the following log-likelihood function:

\begin{equation}
\label{decomp-poly}
-L(d^{\epsilon_1}_i, d^{\epsilon_2}_i) = \sum _{s=1}^{2}  \frac{1}{(\sigma_i^s)^2} {\left(\hat{V}_{i,\mathrm{sca}}^{s}(d^{\epsilon_1}_i, d^{\epsilon_2}_i) - {V}_{i,\mathrm{sca}}^{s} \right)}^{2} ,
\end{equation} 
where $(\sigma_i^s)^2$ is the variance of the extracted  visibility for pixel $i$ and spectral measurement $s$. The variance of the visibility is proportional to the total number of photon counts for the corresponding stepping curve measurement \cite{Mechlem2020_2}.

\subsection{Experimental Measurements}
\begin{figure*}[t!!]
    \centering
    \includegraphics[width=\textwidth]{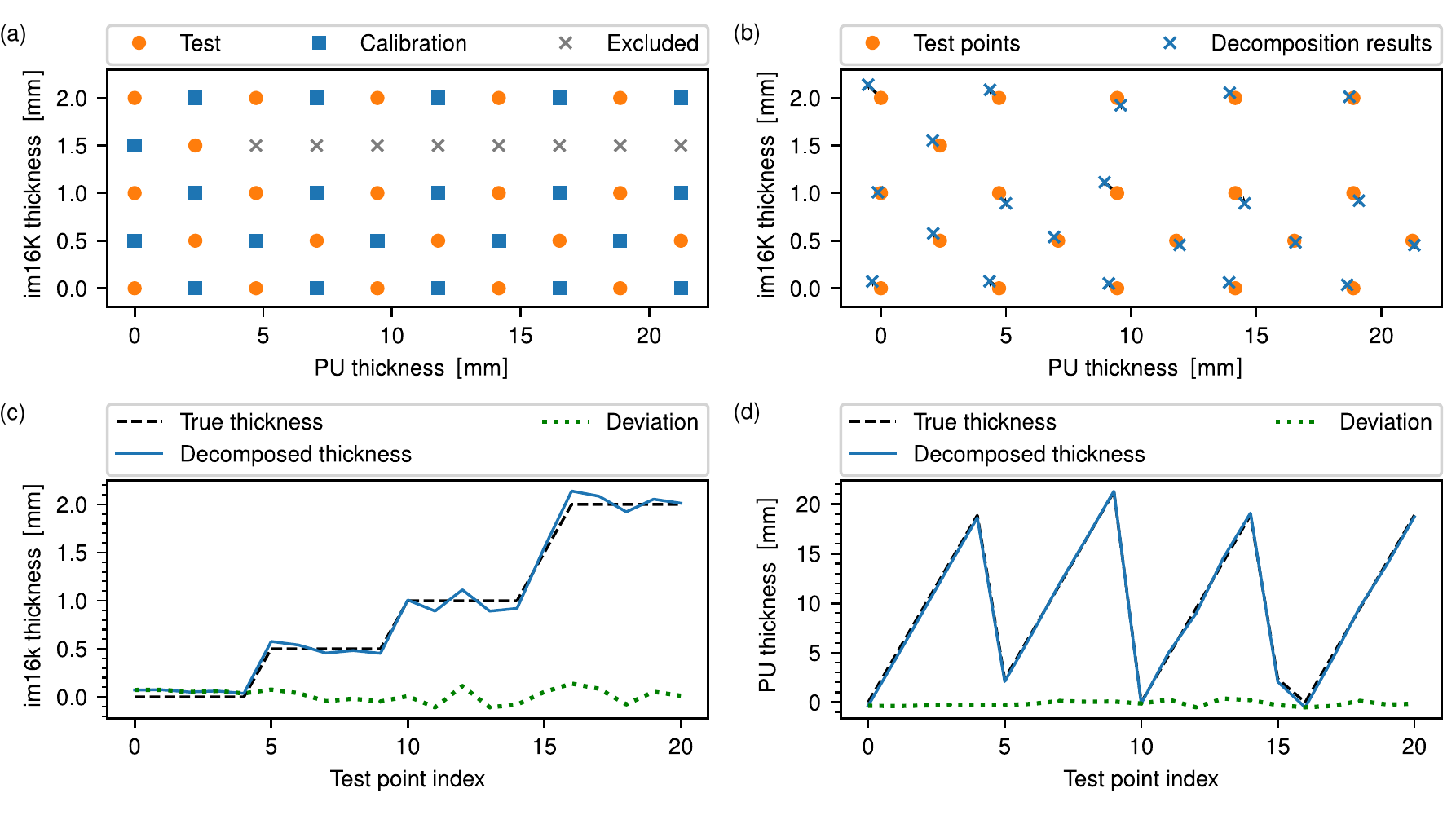}
    \caption{Decomposition accuracy. The quantitative accuracy of the decomposed line integral values was evaluated using a test grid \textbf{(a)} of different thickness combinations that were left out for the calibration of the empirical forward model. The thickness combinations marked with a cross were excluded from both the calibration and test grid due to intensity fluctuations of the X-ray source. The average decomposed thicknesses  are shown in \textbf{(b)} and the deviations from the corresponding ground truth values are plotted for im16K \textbf{(c)} and PU \textbf{(d)}, respectively, where the test point index traverses the test-grid \textbf{(a)} from left to right and then from bottom to top.}
    \label{fig:decomp_acc}
\end{figure*}
\subsubsection{Experimental set-up}
The experimental measurements were performed at a stationary, lab-bench CT setup consisting of an X-ray tube, a flat-panel detector and several positioning devices. The statically mounted X-ray source (XWT-160-CT, X-RayWorX, Garbsen, Germany) is a micro-focus tube with a tungsten reflection target and a $2\,\text{mm}$ thick beryllium window. The data acquisition was performed with a flat-panel detector (PaxScan4030CB, Varex Imaging, Salt Lake City, Utah)  with a $600\,\upmu\text{m}$ thick caesium iodide (CsI) scintillator and an active area of $2048\text{x}1536$ pixels. The detector pixels have a native pixel size of $194\,\upmu\text{m}$ resulting in a maximum field-of-view of $40\text{x}30\,\text{cm}^2$. The specifications of the grating interferometer are listed in table~\ref{specs}. The phantoms used for the calibration of the forward-model were mounted on linear stages between the $G1$ and $G2$ gratings, at a distance of $46.25\,\mathrm{cm}$ in front of the $G2$ grating. In order to determine the energy-dependency of the dark-field signal induced by different materials, an X-ray spectrometer (Amptek X-123CdTe , AMETEK Inc., Berwyn, Pennsylvania, USA) was placed at the position of the flat-panel detector. The spectrometer has a cadmium telluride (CdTe) sensor with a thickness of $1\,\text{mm}$ and an active are of $25\,\text{mm}^2$.
A graphical overview of the experimental set-up is displayed in fig.~\ref{setup}. 
\subsubsection{Sample preparation}
In a first experiment, we performed spectroscopic measurements of two materials with a differently sized microstructure: polyurethane (PU) foam (PORON 4701-30; manufacturer: Rogers Corporation, Chandler, Arizona, USA) and hollow glass micro-bubbles (im16K; manufacturer: 3M, Neuss, Germany). 
The cell size range of the PU foam and the average diameter of the micro-bubbles as provided by the manufacturer are listed in table~\ref{mats}. After evaluation of the variation in the energy-dependent dark-field signal, the two materials were prepared for the calibration of the proposed forward-model (cf. eq.~\eqref{poly}). The polyurethan foam was cut into sheets and stacked to five different thicknesses ($0$ - $2.12\,\text{cm}$). The different stacks were then attached to an aluminum frame which was mounted onto a linear-stage. The im16K powder was filled into five cuboid  plastic containers with different thicknesses ($0$ - $0.2\,\text{cm}$) and attached to a second aluminum frame mounted onto another linear-stage. This arrangement of two movable calibration phantoms allows to measure various thickness combinations of the two materials.\\
For the last experiment, we designed an imaging phantom consisting of the two basis materials PU and im16K. We placed several cylindrical holes in a stack of four PU sheets (with a total thickness of $9.44\,\text{mm}$) to from the letters ``TUM" and filled the holes in the PU sheet at the bottom with im16K powder. The extracted PU cylinders were subsequently placed onto the stack of PU sheets at random locations. Moreover, we added additional cylindrical holes in the topmost sheet of PU without filling them with the im16K powder. The last two steps ensure that the two materials cannot be unambiguously  distinguished in the conventional dark-field images. 
The imaging phantom was placed at the same location as the calibration phantoms in order to ensure the same correlation lengths (see eq. ~\eqref{corr-length}) for all measurements.
\subsubsection{Acquisition parameters and signal extraction}
In order to determine the energy-dependency of the dark-field signal for the two basis materials, we performed a phase stepping with and without the samples in the beam path. The two samples were a $10\,\mathrm{mm}$ thick sheet of PU and a plastic container with a thickness of $2.5\,\mathrm{mm}$ that was filled with im16K powder. The spectrometer was used to measure the  photon spectrum behind the  $G2$ grating for each stepping position. The X-ray tube was operated at $140\,\text{kVp}$ peak voltage. The source power was set to a rather low value of $2\,\text{W}$ to keep the dead-time of the spectrometer within the range recommended by the manufacturer.
Afterwards, we performed phase retrieval \cite{Marschner2016} for each energy bin of the spectrometer separately. 
The process was repeated three times to provide sufficient statistical significance of the acquired data.\\
For the calibration and sample measurement with the flat-panel detector, we generated two different photon spectra by adjusting the tube voltage (to $50 \ \mathrm{kVp}$ and $80 \ \mathrm{kVp}$) and filtering the high energy spectrum with a molybdenum (Mo) foil (cf. fig.~\ref{fig:e-dep-df}(c)). We performed a phase stepping with 7 equidistantly distributed stepping positions for all calibration points and the sample measurement, followed by a pixel-wise signal extraction for both photon spectra. The corresponding acquisition parameters are listed in table~\ref{specs}.

\section{Results}
\subsection{Energy-dependency of the dark-field signal}
We determined the energy-dependency of the linear diffusion coefficients $\epsilon(E)$ for the two basis materials as (compare eq.~\eqref{DF-decomp}):
\begin{equation}
    \epsilon(E) = - \ln {\left( {V_{\mathrm{sca}}(E)}/{V(E)}\right)},
\end{equation}
where $V_{\mathrm{sca}}$ and $V(E)$ represent the visibility spectra that were extracted from the spectroscopic measurements with- and without the sample in the beam path, respectively. fig.~\ref{fig:e-dep-df}a exemplarily shows the measured visibility spectra that were used to determine $\epsilon(E)$ for the PU sample.\\
Grounded on the considerations presented in section~\ref{Spectral Phase-contrast and Darkfield Model} (cf. eq ~\eqref{limit21} and eq.~\eqref{limit22}), we fitted a power-law dependency of the form $\epsilon(E) = cE^{-x}$ to the calculated linear diffusion coefficients that are displayed in fig.~\ref{fig:e-dep-df}(b).
Due to the high transmittance of the absorption gratings around the K-edge of gold, the reference visibility is strongly reduced around $80\,\text{keV}$ which results in a poor signal-to-noise ratio of the extracted data. Therefore, we restricted the fitting range from $30$ to $60\,\text{keV}$. The obtained fit coefficient for the exponent $x$ are listed in table~\ref{mats}. The coefficient $c$ does not affect to the energy-dependency as its only a scaling factor depending on the general signal strength and the sample thickness.
\begin{figure}[t!]
    \centering
    \includegraphics[width=3.5in]{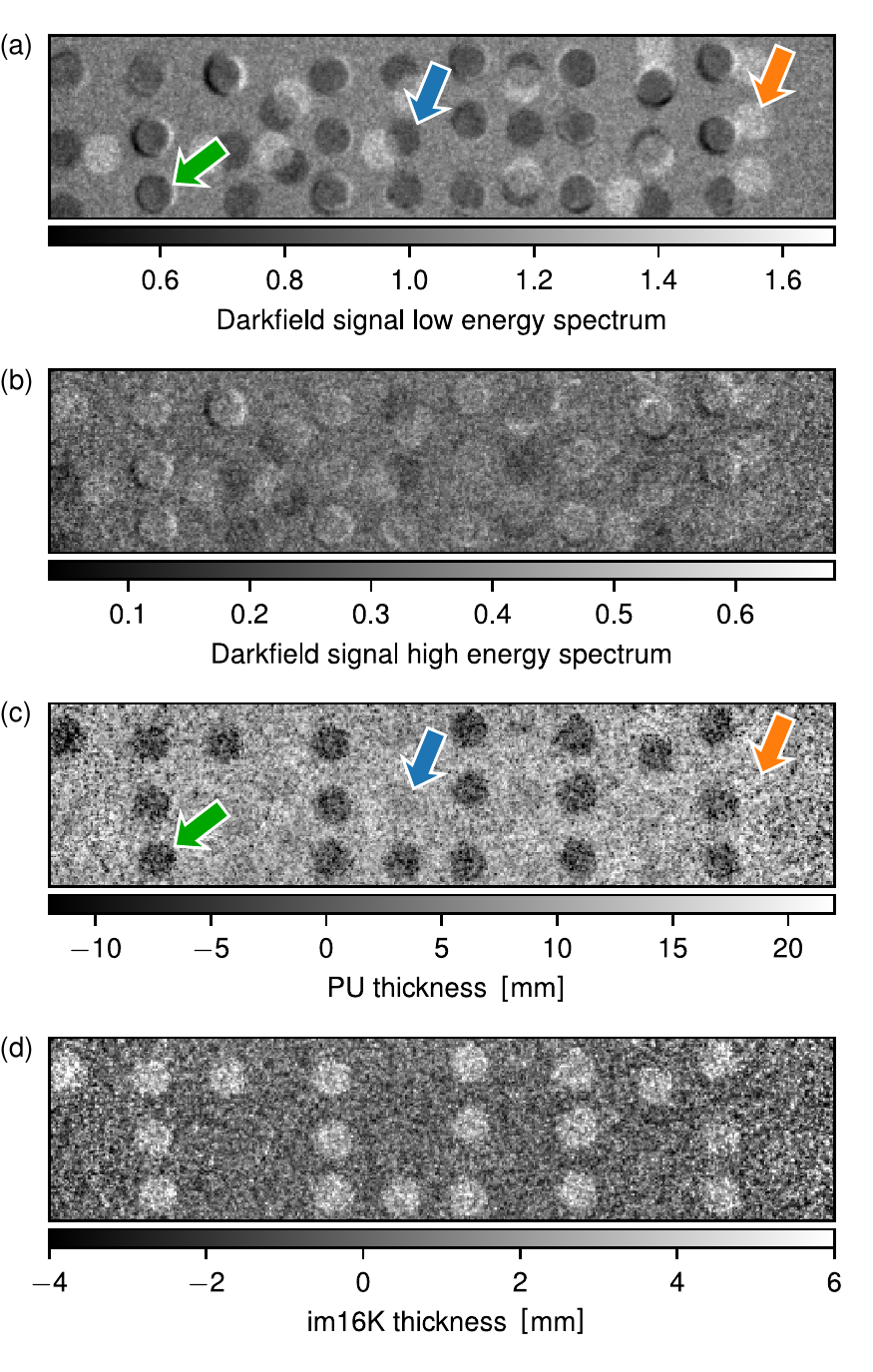}
    \caption{Sample decomposition. In the conventional dark-field signal acquired with the low \textbf{(a)} and high \textbf{(b)} energy spectrum, the PU and im16K structures can not be distinguished properly due to the diminishing contrast. The basis material images, however, provide a clear separation between PU \textbf{(c)} and im16K \textbf{(d)}. Further details on the test phantom are given in the text.}
    \label{fig:decomp_sample}
\end{figure}
\subsection{Decomposition accuracy}
To evaluate the quantitative accuracy of the proposed empirical dark-field material decomposition algorithm, we performed a series of dual-energy dark-field measurements. By measuring all possible thickness combinations of the two calibration phantoms, we obtained 25 measurements of the dark-signal for both the low and high energy spectrum.\\
fig.~\ref{fig:e-dep-df}(d) shows the average low and high energy dark-field log-signals ($-\ln{\left(V^s_{\mathrm{sca}} / V^s \right)}$). Although the dark-field measurements were performed with two well separated X-ray spectra (see fig.~\ref{fig:e-dep-df}(c)), the measured log-signals lie almost on a straight line. This means that the ratio of the low and high energy log-signals varies only slightly for the two basis materials, which is explained by the similar energy dependency of the corresponding linear diffusion coefficients ($\epsilon(E) \propto E^{-3.89}$ and $\epsilon(E) \propto E^{-2.70}$ for PU and im16K, respectively). For comparison purposes, fig.~\ref{fig:e-dep-df}(d) also shows the simulated log-signals for the attenuation channel that would have been obtained if the same calibration measurement was performed with bone and soft tissue as basis materials. In this case, the ratio of the log-signals varies more strongly. This can be explained by the larger differences between the energy-dependent attenuation coefficients $\mu(E)$ compared with the linear diffusion coefficients $\epsilon(E)$. \\
As illustrated in fig.~\ref{fig:decomp_acc}(a), half of the points were used to calibrate the empirical forward-model (cf. eq.~\eqref{cal-poly}), while the remaining data points were decomposed into basis material line integrals. The points marked with crosses had to be excluded from the evaluation as intensity fluctuations of the X-ray tube during the measurement corrupted the acquired data points.
Since the noise level  turned out to be too high to reliably determine the fit coefficients $\vec{c\,}^{s}_{i}$  of the empirical forward model (eq.~\eqref{fwd-emp}) individually for each pixel, we averaged the extracted visibilities $\hat{V}_{i,\mathrm{sca}}^s$ and fitted a common rational function $P^s \left(d_i^{\epsilon_1},d_i^{\epsilon_2}; \vec{c\,}^s \right)$  (see eq.~\eqref{fwd-emp}) for all detector pixel. However, the decomposition into basis material thicknesses according to eq.~\eqref{decomp-poly} was performed individually for each pixel.
The mean of the decomposed thickness values over all detector pixels is shown as a scatterplot in fig.~\ref{fig:decomp_acc}(b) alongside the corresponding ground truth values.  Figures~\ref{fig:decomp_acc}(c) and \ref{fig:decomp_acc}(d) show lineplots of the mean decomposed PU and im16K thicknesses as a function of the test point index together with the corresponding ground truth values. Moreover, the average deviations from the true thicknesses are displayed. 
The average deviation of the decomposed thicknesses values is given by:

\begin{equation}
\label{decomp_mean}
\overline{\Delta d^{\epsilon_{j}}_k} = \frac{1}{M} \sum _{i=1}^{M}  (d^{\epsilon_{j}}_{ik} -D^{\epsilon_{j}}_{k}), \ j \in (1,2),
\end{equation}
where $M$ is the number of detector pixels and $d^{\epsilon_{j}}_{ik}$ is the decomposed thickness value for the  $j$-th dark-field basis material, pixel $i$ and calibration point $k$. The corresponding true thickness values are denoted by $D^{\epsilon_{j}}_{k}$.\\
Although the decomposed thickness values qualitatively match the corresponding ground truth values, quantitative deviations can be observed for all test points. The observed deviations range from $-0.11\,\text{mm}$ to $0.14\,\text{mm}$ for im16K and $-0.51\,\text{mm}$ to $0.38\,\text{mm}$ for PU, where the relative error is higher for im16K than for PU. For both materials the error is relatively constant over all test points and only slightly depends on the corresponding line integral thicknesses. Furthermore the observed deviations exhibit an anti-correlated behavior between the two materials (cf. fig.~\ref{fig:decomp_acc}(b)).

\subsection{Sample decomposition}
Figures ~\ref{fig:decomp_sample}(a) and (b) show the conventional dark-field images of the imaging phantom for the low and high energy spectrum, respectively. The green arrow in figure ~\ref{fig:decomp_sample}(a) exemplarily highlights the location of the holes that were filled with im16K powder to form the letters ``TUM", whereas the blue and orange arrows highlight empty holes and additional PU cylinders that were placed on top of the PU sheets, respectively.
Due to the diminishing contrast, the pattern resulting from the holes filled with im16K powder can hardly be distinguished from the other structures in the two conventional dark-field images in fig.~\ref{fig:decomp_sample}(a,b). Figures~\ref{fig:decomp_sample}(c) and ~\ref{fig:decomp_sample}(d) show the decomposed basis material images of PU and im16K, respectively. Despite the strongly increased noise levels, the basis material images provide a clear separation of PU and im16K revealing the letters ``TUM". Due to the high noise level the holes through just one slice of PU and the additional PU structures on top of the phantom can hardly be distinguished from the background (cf. blue and orange arrow in fig.~\ref{fig:decomp_sample}(d)).

\section{Discussion}

The spectroscopic measurements of the  dark-field signal for PU and im16K demonstrate that the microstructure of the object influences the energy-dependency of the dark-field signal. The observations are in agreement with eq.~\eqref{limit21} and eq.~\eqref{limit22}, which predict that the linear diffusion coefficient $\epsilon(E)$ falls off more quickly for larger microstructures. 
Although eq.~\eqref{limit21} and eq.~\eqref{limit22} are only valid for a diluted suspension of microspheres, the general trends (cf. eq.~\eqref{limit31}- \eqref{limit33}) also apply to more complex microstructures. 
The differences in the energy-dependencies of the dark-field signals induced by the two basis materials PU and im16K are smaller compared to attenuation-based imaging. Consequently,  there is also a smaller difference in the ratio of the log-signals measured for the two dark-field basis materials (cf. green crosses in fig.~\ref{fig:e-dep-df}(c)), even for the two well separated photon spectra that were used in the dual-energy dark-field measurements. This explains the rather low signal-to-noise ratio (SNR) of the dark-field basis material images in fig.~\ref{fig:decomp_sample}. Noise amplification and a degradation of the SNR during material decomposition are well-known problems for attenuation-based dual-energy imaging. Due to the mathematical similarity between dark-field and attenuation-based material decomposition, the same behavior can be expected for dual-energy dark-field imaging. However, the smaller differences in the energy-dependent dark-field signal aggravate the degradation of the SNR. The mathematical similarities between attenuation-based and dark-field dual-energy imaging 
suggest a high efficiency of denoising algorithms, which have originally been developed for attenuation-based dual-energy data \cite{Kalender1988, Mechlem2018}. In the future, we plan to analyze the noise characteristics of dual-energy dark-field imaging in more detail and investigate the viability of existing dual-energy denoising techniques. \\
The evaluation of the decomposition accuracy (cf. fig. \ref{fig:decomp_acc}) shows that quantitative thickness values can be retrieved reliably utilizing the calibration-based dark-field material decomposition approach.
However, the achieved quantitative accuracy is inferior to results for attenuation-based material decomposition methods relying on  empirical forward models \cite{Cardinal1990, Sellerer2019}. In our experiments, this can be attributed at least partly to inaccuracies of the line integral values  of the calibration and test-grid. As could be seen in the conventional X-ray images, the filling density of the im16k powder in the plastic containers fluctuates slightly. This leads to slightly inconsistent input data for the calibration as well as the decomposition of the test-grid. This assumption is supported by the tendency of im16K to have higher deviations for the decomposition results compared to PU, which consists of homogeneous sheets with well-defined thicknesses. \\
The measurement of the imaging phantom (cf. fig.~\ref{fig:decomp_sample}) demonstrates that contributions of different materials to the dark-field signal can be clearly separated and that spatial information about the distribution of different microstructures can be obtained. Moreover, similar to attenuation-based dual-energy imaging, the basis material images provide additional quantitative information compared to the conventional dark-field images. For radiography applications, quantitative line integral values are obtained, whereas an extension of the method to computed tomography imaging would allow to quantify the volume fractions of different mircostructured materials. \\
We believe that the proposed concept of dual-energy dark-field material decomposition could enhance the current performance and extend the capabilities of grating-based dark-field imaging. Especially dark-field chest radiography might benefit from the material decomposition approach.
Dark-field chest radiography potentially allows staging the severeness of pathological changes of the lung tissue with a relatively low radiation dose (compared to chest CT), but the information on the type of the underlying pathological changes is still limited. The combination of dark-field chest radiography with the proposed material decomposition approach might provide valuable quantitative information enabling a more differentiated diagnosis. \\
The presented results demonstrate a proof-of-concept demonstration of dual-energy dark-field material decomposition. The performed experiments involve simplifications that do not fully reflect the conditions of a real diagnostic application. Most importantly, the attenuation caused by the scattering materials evaluated in the study (cf. fig.~\ref{fig:att_vs_df}(d)) is very low and the impact of beam-hardening can be neglected. 
However, especially for larger objects, beam-hardening would alter the measured dark-field signals depending on the absorber thickness in the beam path. In this case, a calibration only relying on the scattering basis materials would be insufficient. In order to take the impact of beam-hardening into account, one would either have to use the full polychromatic forward model given in eq.~\eqref{sdpcdf} or involve equivalent absorbers in the calibration of the empirical forward model. Using the full polychromatic forward-model requires the knowledge of the effective spectrum and the visibility spectrum, which are challenging to  determine with high accuracy. 
An extension of the presented empirical approach would drastically increase the effort for the calibration process. For each dark-field calibration point, the measurement of various equivalent absorber thicknesses would be required. Both approaches should be evaluated with respect to their practical feasibility in future studies. 

\section{Conclusion}
In this work, we introduced the concept of dual-energy X-ray dark-field material decomposition and experimentally demonstrated the method's feasibility. 
Our proof-of-concept study shows that quantitative dark-field basis material line integral values can be obtained  by exploiting differences in the energy-dependent dark-field signals. 
This indicates the potential of dark-field materials decomposition to become a powerful tool for the quantification and differentiation of microstructures within an object. In future applications, this information could be used, for example, to characterize structural changes in the lung parenchyma and thereby allow a more differentiated diagnosis of lung diseases in chest radiography. The applicability of the method for clinical diagnosis should be investigated in future studies.

\bibliographystyle{IEEEtran}
\bibliography{references}

\end{document}